# Magnetoresistance and Scaling Laws in Type-II Weyl Semimetal WP$_2$

V. Nagpal[1], K. S. Jat[2] and S. Patnaik[1,*]

[1]*School of Physical Sciences, Jawaharlal Nehru University, New Delhi, India-110067*
[2]*Government SSNU College, Garoth, Madhya Pradesh, India-458880*

[*]*Corresponding author Email: spatnaik@mail.jnu.ac.in*

**Abstract**

Topological materials with extremely large magnetoresistance exhibit a prognostic feature of resistivity turn-on behaviour. This occurs when the temperature dependence of resistivity $\rho(T)$ changes from metallic to semiconducting characteristics on application of external magnetic field above a threshold value. Here, we study the magneto-transport properties of type-II Weyl Semimetal WP$_2$. The zero field electrical resistivity in the low temperature region indicates the dominant electron-phonon scattering. The saturation in $\rho(T)$ curves under all applied magnetic fields are observed at low temperatures. A minimum in resistivity at ~ 40K is revealed in the temperature derivative of resistivity curves, which implies onset of field induced turn-on effect. Furthermore, a non-saturating linear magnetoresistance (MR) is observed above ~ 5 Tesla which is generally assigned to linear energy dispersion near the Weyl nodes. However, Kohler's scaling fits the data well with resistivity $\rho \sim (B/\rho_0)^m$ that implicitly explains the turn-on behaviour and the resistivity minimum. Thus, semi-classical theories of magnetoresistance are consistent with our data without the need to invoke topological surface states. Our findings in this work provides an alternative basis to understand the temperature dependence of magnetoresistance in topological materials.

## 1. INTRODUCTION

The quest for novel materials exhibiting large magnetoresistance (MR) has evinced much interest in the recent past due to several applications such as magnetic sensors [1], magnetic storage drives [2,3], and magnetic valves [4]. So far, the exploitation of giant magnetoresistance (GMR) in multilayers [5,6] and colossal magnetoresistance (CMR) in manganites [7,8] have provided numerous breakthroughs in the search of extremely large magnetoresistance (XMR) materials. Starting with elemental Bi [9], and graphite [10], in the recent years, such intriguing phenomena has also been reported in several topological materials such as Dirac semimetals $Na_3Bi$ [11,12] and $Cd_3As_2$ [13,14], Weyl semimetals TaAs and NbP [15-18], nodal line semimetals [19], type-II Weyl semimetals such as $WTe_2$ [20] and $MoP_2$ [21,22], transition metal dipnictides $MPn_2$ (M = Nb, Ta; Pn = P, As, Sb) [23-29], and rock salt structured LnX (Ln = La, Nd, Y, Ce; X = Bi, Sb) [30-34].

Despite the fact that XMR is commonly found in these diverse materials, the mechanism of such phenomenon is, however, not yet thoroughly understood. Generally, it is known that MR manifests itself from the charge carrier dynamics beneath the topology of the Fermi surface. Some mechanisms have been proposed for explaining XMR in topologically trivial or non-trivial non-magnetic semimetals [35,36]. The electron-hole carrier compensation, anticipated on the basis of classical two band model, suggests a quadratic MR that is invoked to describe the non-saturating MR in $WTe_2$ [20]. On the other hand, a small difference of electron and hole densities can cause the saturation of MR in the high magnetic field region in compensated metals such as in Bi [9] and graphite [10]. Moreover, in generalized theories, a linear field dependent MR is also predicted in topological semimetals which is ascribed to the linear energy dispersion at the discrete band touching points of the valence and conduction bands in the quantum limit of magnetic field [37,38]. Indeed, the temperature effects in electronic band structure (as well as in the spin-orbit coupled spin textures) detected through angle-resolved photoemission spectroscopy (ARPES) measurements have consistent with the MR measurements of $WTe_2$. In such cases the XMR behaviour shows quadratic field dependence at low magnetic fields, thus favouring the charge compensation mechanism [39]. In contrast, ARPES experimental results on $MoTe_2$ reveals unequal size distributions of electron and hole pockets leading to failure of compensation theory to explain the non-saturating behaviour of XMR in $MoTe_2$ [40]. Similarly, the mobility mismatch between electrons and holes in nearly-perfect compensated semimetal LaSb is suggested to be the reason for its small magnitude of MR [41]. In contrast, in YSb which is devoid of perfect compensation and does not enjoy topological protection, XMR is experimentally observed [42]. ARPES results on MoAs$_2$ imply that the dominant open-orbit topology of its Fermi surfaces rather than closed orbits play important role towards the origin of its quadratic XMR [43]. It is reported that the closed Fermi surface causes the MR to follow the quadratic dependence in weak magnetic fields followed by saturation at high magnetic fields [44]. With regard to WP$_2$, the large non-saturating parabolic magnetoresistance is proposed to originate from the large residual resistivity value, charge compensation and ultrahigh charge carrier mobility [21]. Besides, it is also suggested that the multiband transport and spin textures in WP$_2$ due to a strong spin-orbit coupling might contribute to a huge magnetoresistance [45]. In summary, different mechanisms for XMR is attributed for different compounds, and an interconnection between MR observations and theoretical models pertaining to topological semimetals is necessary.

In this report, we revisit the magneto-transport measurements in a non-centrosymmetric Type-II Weyl semimetal WP$_2$ [46]. In particular, a Type-II Weyl semimetal has Weyl points appearing as contacts between the electron and hole pockets in the band structure. The consequent massless fermions around the Weyl nodes are known as Weyl fermions. The electronic band structure calculations on WP$_2$ exhibit four pairs of type-II Weyl points in which the neighbouring Weyl points possess the same chirality resulting in long topological Fermi arcs on the surface [46]. From our magneto-transport measurements we have found the resistivity plateau at low temperatures and the magnetic field induced resistivity minimum under high magnetic field (11T) around 40K. Further, the transverse MR exhibits a linear non-saturating behaviour at all temperatures. At low temperatures, MR in our sample seems to have low field quadratic as well as high field linear dependence of magnetic field. The low temperature resistivity saturation and field induced resistivity minimum in WP$_2$ can be explained through semi-classical Kohler's scaling law.

## 2. EXPERIMENTAL DETAILS

The polycrystalline sample of WP$_2$ was successfully synthesized by the solid state phosphating reaction method [47]. The raw materials tungsten trioxide (WO$_3$) and red phosphorus powders in respective amounts of 500mg and 300mg were uniformly mixed to a fine ground powder. The ground mixture was then sealed in an evacuated silica tube. The tube, kept in a programmable furnace, was then heated to 950°C at a rate of 10°C/min and dwell for 5 hours at this temperature. The tube after naturally cooled to room temperature was taken out from the furnace. The reaction products containing black coloured powders along with a few white powders were treated with water/acid several times to obtain the final product. The crystal structure of the resultant synthesized sample was examined using the powder X-ray diffractometer (Rigaku/DMAX 2500PC). The Reitveld refined X-ray diffraction (XRD) pattern of powdered WP$_2$ sample is shown in Fig. 1. The diffraction pattern clearly mentions the single phase of the as-synthesized compound, confirming that WP$_2$ crystallizes in orthorhombic structure with spacegroup $Cmc2_1$ (see the right inset of Fig. 1). The different refinement parameters such as $\chi^2 = 2.65$, $R_{wp} = 18.3$, $R_p = 16.2$, $R_{exp} = 11.26$ implies the goodness of Reitveld refined data. The refined XRD data obtains the lattice parameters: $a = 3.159$ Å, $b = 11.16$Å and $c = 4.984$Å, consistent with the band structure calculations [46]. Moreover, the elemental composition of polycrystalline WP$_2$ is determined from Energy Dispersive X-ray (EDX) analysis. It has been found that the elemental ratio i.e., W:P = 1:1.86 is close to the stoichiometric WP$_2$, as illustrated in the left inset of Fig. 1. The longitudinal electrical resistivity and magnetoresistance measurements on WP$_2$ sample were carried in 8T Cryogenic Free Magnet (CFM) system equipped with the lowest temperature at 1.6K. The electrical resistivity at a high magnetic field H was performed out in Physical Properties Measurement System (PPMS). The standard four probe method was employed for the resistivity measurement.

## 3. RESULTS AND DISCUSSIONS

Fig. 2(a) shows the temperature dependence of longitudinal resistivity of WP$_2$ from 2-300K in the absence of applied magnetic field. The resistivity $\rho(2K)$ at 2K is found to be 4.2μΩ-cm that decreases monotonically with the temperature, exhibiting the metallic behaviour of sample. The residual resistivity in polycrystalline WP$_2$ is comparatively larger than that reported in WP2 single crystals.[21,45] This high comparable value in polycrystalline WP$_2$ is due to the grain boundary effect where the charge blocking layers are formed near the grain boundaries, thus scattering the carriers that results in low conductivity and carrier mobility. The calculated residual resistivity ratio $[RRR = \rho(290K)/\rho(2K)]$ is 22. This is a very high value considering that fact that WP$_2$ is not a layered compound and the sample is in polycrystalline form. The zero field resistivity at low temperatures (Fig.2 (a)) is fitted with equation: $\rho_{xx}(T) = \rho_{xx}(0) + BT^n$, where $\rho_{xx}(0)$ and $B$ are the residual resistivity and a constant factor, respectively. The exponent $n$ value reflects the dominant scattering mechanism followed in the system. From the fitting, the derived parameters are $\rho_{xx}(0) = 3.97 \mu\Omega cm$, $B = 3.28 \times 10^{-11} \Omega cm K^{-n}$ and $n = 2.8$. The fitted exponent value is obtained close to the value $n = 3$, which implies the dominance of electron-phonon mechanism below 90K in the sample. The otherwise limiting $n = 2$ value means the prominent electron-electron scattering in the system. The inset in Fig. 2(a) depicts the low temperature resistivity $\rho$ versus $T^3$ at zero applied magnetic field where a linearity in plot is observed at temperatures below 90K. Moreover, the resistivity at zero applied magnetic field in the range 2-100K is fitted with an equation: $\rho(T) = \rho_0 + aT^2 + bT^5 + c^*exp(-T_0/T)$, as shown in Fig. 2(b) (red colour). Here, $\rho_0, a, b$ and $T_0$ are predetermined constant fitting factors. The terms $T^2$ and $T^5$ define the electron-electron and electron-phonon interactions, respectively while the exponential term is due to the phonon drag mechanism due to which the resistivity shows a steep downward curve. The $T^2$ and $T^5$ scattering terms do not separately fit to the resistivity data in the measured range and hence, a combination with phonon drag term is introduced so as to fit the resistivity. Also, the zero field resistivity in the high temperature region above 250K is observed to be linear with respect to the temperature as illustrated in Fig. 2(b) (green colour) which implies dominant electron-phonon interactions in WP$_2$.

Next, the resistivity as a function of temperature under different applied magnetic fields measured in the temperature range 2-150K is depicted in Fig. 3(a). The applied magnetic field *B* is oriented perpendicular to the current *I*. It is highlighted that the resistivity at fields ($\leq 7T$) in the whole temperature range remains metallic and tend to increases rapidly in the low temperature region upon the application of magnetic field. However, we see a dramatic change in resistivity behaviour at 11T applied field where the resistivity first decreases with the temperature and then increases further below a certain temperature $T^*$, indicating a minimum in resistivity, marked as an inflection point. In addition, the resistivity at all applied fields above 100K does not enhanced much with the increasing applied magnetic field, unlike the behaviour observed at low temperatures. Further, the inset in Fig. 3(a) illustrates the enlarged view of resistivity in the low temperature regime. At 11T field, a resistivity minimum is observed at a temperature $T^*$ around 40K beyond which the resistivity seems to increase dramatically. Similar magnetic field driven resistivity turn-on behaviour has been reported even at much lower applied magnetic fields in earlier results of WP$_2$ single crystals [21,45] and other known Weyl semimetals [20,22,48]. Such type of resistivity crossover is allocated to metal-insulator transition (MIT) proposed to originate due to the opening of a bulk gap at the band crossing points [49]. Concurrently, some scaling analysis has also been suggested that could explain the MIT [50,51]. It is also observed that the zero field resistivity and resistivity at all magnetic fields saturates below 30K. The low temperature resistivity saturation scenario observed in several other reported topological materials might arise due to the contribution of 2D surface states conduction dominant over the bulk transport conduction or some kind of symmetry breaking in a material [52-54].

Furthermore, the derivative of resistivity $d\rho/dT$ vs. temperature for different magnetic fields is plotted in Fig. 3(b). Strikingly, the resultant curve for 11T applied field exhibits two distinct characteristic temperatures i.e., the resistivity crossover temperature $T_m$, where a sign change is observed in $d\rho/dT$ and the temperature $T_i$ at which the $d\rho/dT$ curve is minimum and below which the resistivity saturation begin to appear. At an applied magnetic field of 11T, the characteristic temperatures $T_m$ and $T_i$ are found to be 39K and 21K, respectively. On the other hand, we do not see any such characteristic minimum in $d\rho/dT$ plot at lower magnetic fields, thereby implying the absence of field induced turn-on behaviour or MIT feature as also evident from the resistivity curve in Fig. 3(a). The magnetic field induced resistivity turn-on behaviour further implies the thermally activated transport. Next, the resistivity at low temperatures is analysed through Arrhenius equation: $\rho = \rho_0 \exp(E_g/k_B T)$ in the plot of $\ln \rho$ vs. $1/T$, as shown in Fig.3(c). The slopes determined through the fit gives the calculated band gap values of 0.014meV and 0.093meV for 7T and 11T applied field, respectively. These gap values are comparatively much lower than previous WP$_2$ results and other topological materials [21,54]. However, the value of band gap is increased from 7T to 11T, which indicates that the metallicity decreases upon the application of magnetic field in the low temperature region. In Fig. 3(d), the temperature dependent magnetoresistance $MR(T) = (\rho(T,B) - \rho(T,0))/\rho(0T) \times 100\%$ calculated at fixed magnetic fields is presented, where $\rho(T,0)$ is the resistivity in the absence of field and $\rho(T,B)$ is the resistivity under fixed applied field. The normalized MR(T) for all fields evaluated as the temperature dependent MR(T) divided by their respective MR(T) values at 2K i.e., MR/MR(2K) is shown in the inset of Fig. 3(d). The normalized MR(T) curves for different magnetic fields merge into a single curve signifying the same temperature dependence of normalized MR(T) for all magnetic fields. This type of behaviour is reported in various topological materials and ascribed to the exclusion of field-induced band gap opening mechanism [48,54-55].

Fig. 4(a) displays the transverse magnetoresistance MR as a function of magnetic field $B$ upto 6T applied field at various constant temperatures. The direction of current $I$ and magnetic field is perpendicular to each other ($I \perp B$). The highest positive MR value of 118% is attained at 2K and 6T, and any sign of saturation is not observed upto the highest applied magnetic field. Our experimental MR value is found reasonably lower than the published results on single crystals of $WP_2$ where MR reaches an extraordinarily large value at 14,300% at 2K and 9T [21,45]. The lower value of MR found in our sample is expected due to the polycrystalline nature of the specimen in which the grain boundary effects play a significant role. This is also inferred from the low carrier mobilities and failure of charge compensation in our sample that lead to small MR [21,45]. It is also clearly seen that a little variation in MR values is observed only at low temperatures between 2K and 5K in the entire field range when compared to the values at other higher temperatures. Further, the transverse MR can be represented by the Kohler scaling rule which explains the mechanisms involved in scattering of charge carriers. In connection with semi-classical band theory, the Kohler rule is governed by the following equation.[56]

$$MR = \frac{\Delta\rho}{\rho_0} = \alpha\left(\frac{B}{\rho_0}\right)^m \qquad (1)$$

Here, the parameters $(\alpha, m)$ in Equation (1) are the predetermined constants and $\rho_0$ is the zero field resistivity. Following this law, the transverse MR data is plotted as $MR \sim B/\rho_0$ curve at all temperatures using the Kohler scaling as shown in Fig. 4(b). It is noted that the MR curves at distinct temperatures do not collapse onto a single curve at temperatures above 5K [45]. According to Kohler's rule, the unification of MR curves to a single line suggests that the charge carries in the system follows the same scattering mechanism at all temperatures. Conversely, the breakdown of Kohler scaling analysis in $WP_2$ might be due to the various factors such as different densities or temperature dependent mobilities, and/or different concentrations of two or more types of charge carriers with different scattering times [57]. Additionally, MR vs $B/\rho_0$ curves at all temperatures have been fitted with Equation (1), as depicted in Fig. 4(c). The fitting to MR curves gives the information about the parameters $\alpha$ and $m$ at respective temperatures. The temperature dependence of the extracted values of fitted parameters are displayed in Fig. 4(d). The parameter $\alpha$ increases with increasing temperature and $m$ decreases with the increase in temperature. It is also found that MR is nearly linear dependent on magnetic field at low temperatures while an exact linear field dependence i.e., $m = 1$ is observed at 150K in $WP_2$. On the contrary, a near quadratic field dependence of MR is found in previous MR results on $WP_2$ [21,45]. A similar linear MR behaviour is often reported in various Dirac and Weyl semimetals [58-60, 22]. The linear field dependent MR in $WP_2$ is ascribed to the role of linearly dispersing Weyl fermions at all temperatures [61].

For better understanding the MR dependence on magnetic field in $WP_2$, the field dependent MR curves measured at 2K and 5K, as depicted in Fig. 5(a) are fitted with a quadratic equation: $MR = a_1 B + a_2 B^2$ where $a_1$ and $a_2$ corresponds to the fitting parameters. This, in turn, indicates that MR at low temperatures in $WP_2$ is due to the contributions of both the linear as well as parabolic terms of magnetic field. The contributory squared field dependence is quite accredited to the conduction electrons deflected under an applied magnetic field due to the Lorentz force, as explained in the semi-classical model. However, the concept of linear MR origin, though controversial, has not been justified till date. The linear MR observed in semiconductors is assigned to the high disorder and strong inhomogeneous phase of the sample [62]. Abrikosov et al theorized that the linear band crossings in the bulk systems could unveil the linear MR in the quantum regime where the charged particles are considered as massless Dirac fermions [38]. In order to analyse the quadratic relation of MR, the field derivative of MR, $dMR/dB$, against the magnetic field at 2K and 5K is presented in Fig. 5(b). It can be viewed clearly that the $dMR/dB$ plot at both the temperatures reveal the linear behaviour over a broad field range which approves the semiclassical $MR \sim a_2 B^2$. Furthermore, a saturation behaviour is observed in $dMR/dB$ curve for both the temperatures above a certain critical magnetic field. This suggests the crossover from $B^2$ behaviour to linear field dependence at a characteristic field $B^*$ whereby the crossover field is determined from the intersection of slopes of the linear and saturating curve, as presented in the Fig. 5(b) inset. The characteristic fields are deduced as 5.3T and 5.6T at 2K and 5K, respectively. This saturating $dMR/dB$ behaviour clearly signifies that the MR predominantly follows the linear field dependence in addition to a small quadratic term i.e., $MR = a_1 B + o(B^2)$ above the characteristic field. It is stated that the orbital quantization of charge carriers with linear energy dispersion under an external applied magnetic field could lead to the quantized Landau levels (LLs) with energies $E_n = \text{sgn}(n)\sqrt{2e\hbar B |n|}$ [63]. So when the splitting, expressed as $\Delta_{LL} = \pm v_F \sqrt{2e\hbar B}$, between the first and lowest Landau level at a particular temperature and the field in the quantum limit becomes large compared to the Fermi energy as well as the thermal fluctuations, the lowest Landau level is occupied by all the charge carriers which behave as Dirac or Weyl fermions in Weyl semimetal and as a result, linear magnetoresistance arises in the quantum transport [63]. Therefore, the linear MR in Weyl semimetal $WP_2$ possibly originates from electron and hole pockets exhibiting the linear energy dispersion near the Weyl nodes [61].

As stated earlier, Kohler rule is originally described by Equation (1) which accounts for the magnetoresistance in materials. Thus, in order to explore how Equation (1) can also lead to astonishing turn-on behaviour, we again replot the resistivities $\rho_{xx}(T, B)$ for zero field and 11T applied field and their difference $\Delta\rho_{xx} = \rho_{xx}(T, 11) - \rho_{xx}(T, 0)$, as displayed in Fig. 6. It is clearly shown that the sample resistivity under an applied magnetic field comprises of two components of resistivity i.e., $\rho_0 (\equiv \rho_{xx}(T, 0))$ and $\Delta\rho_{xx}$. The components $\rho_0$ and $\Delta\rho_{xx}$ have opposite temperature dependencies meaning the first component $\rho_0$ increases whereas the second component $\Delta\rho_{xx}$ decreases with the temperature. So, Equation (1) can further be rewritten in the form:

$$\rho_{xx}(T, B) = \rho_0 + \alpha\, B^m / \rho_0^{m-1} \qquad (2)$$

The zero field resistivity $\rho_0$, being the only temperature dependent term in Equation (2), solely determines the temperature variation of $\rho_{xx}(T, B)$. The second term refers to the magnetic-field induced resistivity $\Delta\rho_{xx}$ following the Kohler's rule and is inversely proportional to $1/\rho_0$ (when $m = 2$). The two terms compete with each other upon varying the temperature, which results in a possible minimum at $T^*$ in $\rho_{xx}(T, B)$ curves. At $T < T^*$, the resistivity at a fixed magnetic field show a dramatic increase as the temperature decreases while at $T > T^*$, a metallic behaviour is observed similar to that of temperature dependence of zero field resistivity. The resistivity $\rho_{xx}(11T)$ and $\Delta\rho_{xx}$ curves in Fig. 5 are well fitted with Equation (2) with $\alpha = 0.565\ (\mu\Omega - cm/T)^{1.54}$ and $m = 1.54$, different from the values extracted from our MR results identified in Fig. 4(d), indicating the precisely unequal densities of charge carriers in our sample [55]. The good fit of the resistivity ensures the occurrence of a resistivity minimum at $T^*$ in $\rho_{xx}(T, B)$ curve beyond the critical magnetic field $B_C$. For convenience, $T^*$ can be derived from Equation (2) by using $\rho_{xx}(T, B)$ i.e., $d\rho_{xx}(T, B)/dT = 0$ which results to an expression $\rho_0(T^*) = B[\alpha(m-1)]^{1/m}$. Hence, the critical field $B_C$ can be determined from the relation: $B_C = \rho_0[\alpha(m-1)]^{-1/m}$. The resistivity $\rho_0 = 3.97\ \mu\Omega - cm$ is derived from the fitting equation $\rho_0 = A + BT^n; n = 3$ shown in Fig. 2(a). Using the values of $\rho_0$, $\alpha$ and $m$, $B_C$ is evaluated to be 8.58T, in consistent with the experimental measured data where absence of minima is clearly noted in the resistivity measurements at 7T and other lower applied fields. This implies that Kohler's rule defined by Equation (1) can estimate the qualitative description of the resistivity minimum and the remarkable turn-on behaviour included in the total resistivity at low temperatures in WP$_2$. Thus the features ascribed to topological surface states from magnetoresistance data are well accounted by a semi-classical theory as well.

## 4. CONCLUSION

In conclusion, we have successfully synthesized the type-II Weyl Semimetal $WP_2$. We have found the field-induced resistivity turn-on behaviour under a high applied magnetic field around 40K followed by the resistivity saturation in the low temperature region for different applied fields. The unexpected low value of transverse magnetoresistance with non-saturating behaviour suggests failure of electron-hole compensation theory in $WP_2$. We have determined the linear field dependence of MR at all the fixed temperatures from the power law, indicating the presence of linear energy dispersing Weyl fermions in $WP_2$. The temperature derivative of transverse MR at low temperatures indicates that the MR follows a semi-classical parabolic field dependence at low fields in addition to linear MR in the high field region. We have shown the resistivity crossover and the resistivity minimum are well described by the Kohler's scaling rule, negating the phenomenon of field-driven metal-insulator transition in $WP_2$.

## ACKNOWLEDGEMENT

V. Nagpal acknowledges UGC NET-SRF Fellowship, UGC, New Delhi for financial support. S. Patnaik thanks funding support of EMR/2016/003998/PHY, DST-FIST, and DST-PURSE programs of the Government of India.

**References**

[1] J. Jankowski, S. El-Ahmar, and M. Oszwaldowski, "Hall Sensors for Extreme Temperatures", *Sensors* **11** (2011) 876–885.

[2] C. Rao and A. Cheetham, "Giant Magnetoresistance in Transition Metal Oxides". *Science* **272** (1996) 369–370.

[3] J. M. Daughton, "GMR applications", *Journal of Magnetism and Magnetic Materials* **192** (1999), 334–342.

[4] S. Wolf, D. Awschalom, R. Buhrman, J. Daughton, S. Von Molnar, M. Roukes, A.Y. Chtchelkanova, and D. Treger, "Spintronics: A Spin-Based Electronics Vision for the Future" *Science* **294** (2001) 1488.

[5] M. N. Baibich, J. M. Broto, A. Fert, F. Nguyen Van Dau, F. Petroff, P. Etienne, G. Creuzet, A. Friederich, and J. Chazelas, "Giant Magnetoresistance of (001)Fe/(001)Cr Magnetic Superlattices", *Physical Review Letters* **61** (1988) 2472–2475.

[6] G. Binasch, P. Grünberg, F. Saurenbach, and W. Zinn, "Enhanced magnetoresistance in layered magnetic structures with antiferromagnetic interlayer exchange". *Physical Review B* **39** (1989) 4828–4830.

[7] Y. Moritomo, A. Asamitsu, H. Kuwahara, and Y. Tokura, "Giant magnetoresistance of manganese oxides with a layered perovskite structure". *Nature* **380** (1996) 141–144.

[8] M. B. Salamon and M. Jaime, "The physics of manganites: Structure and transport". *Reviews of Modern Physics* **73** (2001) 583–628.

[9] B. Abeles and S. Meiboom, "Galvanomagnetic Effects in Bismuth". *Physical Review* **101** (1956) 544–550.

[10] Y. Iye, P. M. Tedrow, G. Timp, M. Shayegan, M. S. Dresselhaus, G. Dresselhaus, A. Furukawa, and S. Tanuma, "High-magnetic-field electronic phase transition in graphite observed by magnetoresistance anomaly". *Physical Review B* **25** (1982) 5478–5485.

[11] Z. Wang, Y. Sun, X.-Q. Chen, C. Franchini, G. Xu, H. Weng, X. Dai, and Z. Fang, "Dirac semimetal and topological phase transitions in $A_3Bi$ (A=Na, K, Rb)" *Phys. Rev. B* **85** (2012) 195320.

[12] J. Xiong, S. K. Kushwaha, T. Liang, J. W. Krizan, M. Hirschberger, W. Wang, R. Cava, and N. Ong, "Evidence for the chiral anomaly in the Dirac semimetal Na$_3$Bi". *Science* **350** (2015) 413–416.

[13] L. P. He, X. C. Hong, J. K. Dong, J. Pan, Z. Zhang, J. Zhang, and S. Y. Li, "Quantum Transport Evidence for the Three-Dimensional Dirac Semimetal Phase in Cd$_3$As$_2$". *Physical Review Letters* **113** (2014) 246402.

[14] T. Liang, Q. Gibson, M. N. Ali, M. Liu, R. Cava, and N. Ong, "Ultrahigh mobility and giant magnetoresistance in the Dirac semimetal Cd$_3$As$_2$". *Nature Materials* **14** (2015) 280–284.

[15] X. Huang, L. Zhao, Y. Long, P. Wang, D. Chen, Z. Yang, H. Liang, M. Xue, H. Weng, Z. Fang et al., "Observation of the Chiral-Anomaly-Induced Negative Magnetoresistance in 3D Weyl Semimetal TaAs", *Physical Rev X* **5** (2015) 031023.

[16] C. Shekhar, A. K. Nayak, Y. Sun, M. Schmidt, M. Nicklas, I. Leermakers, U. Zeitler, Y. Skourski, J. Wosnitza, Z. Liu et al., "Extremely large magnetoresistance and ultrahigh mobility in the topological Weyl semimetal candidate NbP". *Nature Physics* **11** (2015) 645–649.

[17] Y. Luo, N. J. Ghimire, M. Wartenbe, H. Choi, M. Neupane, R. D. McDonald, E. D. Bauer, J. Zhu, J. D. Thompson, and F. Ronning, Electron-hole compensation effect between topologically trivial electrons and nontrivial holes in NbAs", *Phys. Rev. B* **92** (2015) 205134.

[18] J. Du, H. Wang, Q. Chen, Q. Mao, R. Khan, B. Xu, Y. Zhou, Y. Zhang, J. Yang, B. Chen, C. Feng, and M. Fang, "Large unsaturated positive and negative magnetoresistance in Weyl semimetal TaP", *Sci. China: Phys., Mech. Astron.* **59** (2016) 657406.

[19] T. Bzdušek, Q.Wu, A. Rüegg, M. Sigrist, and A. A. Soluyanov, "Nodal-chain metals". *Nature* **538** (2016) 75–78.

[20] M. N. Ali, J. Xiong, S. Flynn, J. Tao, Q. D. Gibson, L. M. Schoop, T. Liang, N. Haldolaarachchige, M. Hirschberger, N. Ong, et al., "Large, non-saturating magnetoresistance in WTe$_2$". *Nature* **514** (2014) 205–208.

[21] N. Kumar, Y. Sun, N. Xu, K. Manna, M. Yao, V. Süss, I. Leermakers, O. Young, T. Förster, M. Schmidt, H. Borrmann, B. Yan, U. Zeitler, M. Shi, C. Felser & C. Shekhar, "Extremely high magnetoresistance and conductivity in the type-II Weyl semimetals WP$_2$ and MoP$_2$". *Nat. Comm.* **8** (2017) 1642.

[12] J. Xiong, S. K. Kushwaha, T. Liang, J. W. Krizan, M. Hirschberger, W. Wang, R. Cava, and N. Ong, "Evidence for the chiral anomaly in the Dirac semimetal Na$_3$Bi". *Science* **350** (2015) 413–416.

[13] L. P. He, X. C. Hong, J. K. Dong, J. Pan, Z. Zhang, J. Zhang, and S. Y. Li, "Quantum Transport Evidence for the Three-Dimensional Dirac Semimetal Phase in Cd$_3$As$_2$". *Physical Review Letters* **113** (2014) 246402.

[14] T. Liang, Q. Gibson, M. N. Ali, M. Liu, R. Cava, and N. Ong, "Ultrahigh mobility and giant magnetoresistance in the Dirac semimetal Cd$_3$As$_2$". *Nature Materials* **14** (2015) 280–284.

[15] X. Huang, L. Zhao, Y. Long, P. Wang, D. Chen, Z. Yang, H. Liang, M. Xue, H. Weng, Z. Fang et al., "Observation of the Chiral-Anomaly-Induced Negative Magnetoresistance in 3D Weyl Semimetal TaAs", *Physical Rev X* **5** (2015) 031023.

[16] C. Shekhar, A. K. Nayak, Y. Sun, M. Schmidt, M. Nicklas, I. Leermakers, U. Zeitler, Y. Skourski, J. Wosnitza, Z. Liu et al., "Extremely large magnetoresistance and ultrahigh mobility in the topological Weyl semimetal candidate NbP". *Nature Physics* **11** (2015) 645–649.

[17] Y. Luo, N. J. Ghimire, M. Wartenbe, H. Choi, M. Neupane, R. D. McDonald, E. D. Bauer, J. Zhu, J. D. Thompson, and F. Ronning, Electron-hole compensation effect between topologically trivial electrons and nontrivial holes in NbAs", *Phys. Rev. B* **92** (2015) 205134.

[18] J. Du, H. Wang, Q. Chen, Q. Mao, R. Khan, B. Xu, Y. Zhou, Y. Zhang, J. Yang, B. Chen, C. Feng, and M. Fang, "Large unsaturated positive and negative magnetoresistance in Weyl semimetal TaP", *Sci. China: Phys., Mech. Astron.* **59** (2016) 657406.

[19] T. Bzdušek, Q.Wu, A. Rüegg, M. Sigrist, and A. A. Soluyanov, "Nodal-chain metals". *Nature* **538** (2016) 75–78.

[20] M. N. Ali, J. Xiong, S. Flynn, J. Tao, Q. D. Gibson, L. M. Schoop, T. Liang, N. Haldolaarachchige, M. Hirschberger, N. Ong, et al., "Large, non-saturating magnetoresistance in WTe$_2$". *Nature* **514** (2014) 205–208.

[21] N. Kumar, Y. Sun, N. Xu, K. Manna, M. Yao, V. Süss, I. Leermakers, O. Young, T. Förster, M. Schmidt, H. Borrmann, B. Yan, U. Zeitler, M. Shi, C. Felser & C. Shekhar, "Extremely high magnetoresistance and conductivity in the type-II Weyl semimetals WP$_2$ and MoP$_2$". *Nat. Comm.* **8** (2017) 1642.

[22] A. Wang, D. Graf, A. Stein, Y. Liu, W. Yin, and C. Petrovic, "Magnetotransport properties of MoP$_2$", *Phys. Rev. B* **96** (2017) 195107.

[23] K. Wang, D. Graf, L. Li, L. Wang, and C. Petrovic, "Anisotropic giant magnetoresistance in NbSb2". *Sci. Rep.* **4** (2015) 7328.

[24] B. Shen, X. Deng, G. Kotliar, and N. Ni, Fermi surface topology and negative longitudinal magnetoresistance observed in the semimetal NbAs$_2$". *Phys. Rev. B* **93** (2016), 195119.

[25] D. Wu, J. Liao, W. Yi, X. Wang, P. Li, H. Weng, Y. Shi, Y. Li, J. Luo, X. Dai et al., "Giant semiclassical magnetoresistance in high mobility TaAs$_2$ semimetal". *Appl. Phys. Lett.* **108** (2016) 042105.

[26] Y. Li, L. Li, J. Wang, T. Wang, X. Xu, C. Xi, C. Cao, and J. Dai, "Resistivity plateau and negative magnetoresistance in the topological semimetal TaSb$_2$", *Phys. Rev. B* **94** (2016) 121115.

[27] Y. Luo, R. McDonald, P. Rosa, B. Scott, N. Wakeham, N. Ghimire, E. Bauer, J. Thompson, and F. Ronning, Anomalous electronic structure and magnetoresistance in TaAs$_2$". *Sci. Rep.* **6** (2016) 27294.

[28] Y.-Y. Wang, Q.-H. Yu, P.-J. Guo, K. Liu, and T.-L. Xia, "Resistivity plateau and extremely large magnetoresistance in NbAs$_2$ and TaAs$_2$", *Phys. Rev. B* **94** (2016) 041103.

[29] Z. Wang, Y. Li, Y. Lu, Z. Shen, F. Sheng, C. Feng, Y. Zheng, and Z. Xu, arXiv:1603.01717.

[30] F. Tafti, Q. Gibson, S. Kushwaha, N. Haldolaarachchige, and R. Cava, "Resistivity plateau and extreme magnetoresistance in LaSb". *Nature Physics* **12** (2016) 272–277.

[31] N. Kumar, C. Shekhar, S.-C. Wu, I. Leermakers, O. Young, U. Zeitler, B. Yan, and C. Felser, "Observation of pseudo-two-dimensional electron transport in the rock salt-type topological semimetal LaBi". *Phys. Rev. B* **93** (2016) 241106.

[32] N. Wakeham, E. D. Bauer, M. Neupane, and F. Ronning, "Large magnetoresistance in the antiferromagnetic semimetal NdSb". *Phys. Rev. B* **93** (2016) 205152.

[33] S. Sun, Q. Wang, P.-J. Guo, K. Liu, and H. Lei, "Large magnetoresistance in LaBi: origin of field-induced resistivity upturn and plateau in compensated semimetals". *New J. Phys*. **18**, (2016) 082002.

[34] O. Pavlosiuk, P. Swatek, and P. Wiśniewski, "Giant magnetoresistance, three-dimensional Fermi surface and origin of resistivity plateau in YSb semimetal". *Sci. Rep.* **6** (2016) 38691.

[35] M. Parish and P. Littlewood, ""Non-saturating magnetoresistance in heavily disordered semiconductors". *Nature* **426** (2003) 162–165.

[36] J. M. Ziman, "Principles of the Theory of Solids". In: Cambridge University Press, 1972.

[37] A. A. Abrikosov, "Quantum magnetoresistance". *Phys. Rev. B* **58** (1998) 2788–2794.

[38] A. Abrikosov, "Quantum linear magnetoresistance". *Europhysics Letters (EPL)* **49**(2000) 789–793.

[39] Y. Wu, N. H. Jo, M. Ochi, L. Huang, D. Mou, S. L. Bud'ko, P. C. Canfield, N. Trivedi, R. Arita, and A. Kaminski, "Temperature-Induced Lifshitz Transition in $WTe_2$". *Phys. Rev. Lett.* **115** (2015) 166602.

[40] S. Thirupathaiah, R. Jha, B. Pal, J. S. Matias, P. K. Das, P. K. Sivakumar, I. Vobornik, N. C. Plumb, M. Shi, R. A. Ribeiro, and D. D. Sarma, "$MoTe_2$: An uncompensated semimetal with extremely large magnetoresistance". *Phys. Rev. B* **95** (2017) 241105.

[41] H. Y. Yang, T. Nummy, H. Li, S. Jaszewski, M. Abramchuk, D. S. Dessau, and F. Tafti, *Phys. Rev. B* **96**, 235128 (2017).

[42] J. He. et. al., "Distinct Electronic Structure for the Extreme Magnetoresistance in YSb". *Phys. Rev. Lett.* **117** (2016) 267201.

[43] R. Lou, Y. Xu, L.-X. Zhao, Z.-Q. Han, P.-J. Guo, M. Li, J.-C. Wang, B.-B. Fu, Z.-H. Liu, Y.-B. Huang et al., "Observation of open-orbit Fermi surface topology in the extremely large magnetoresistance semimetal $MoAs_2$". *Phys. Rev. B* **96** (2017) 241106.

[44] S. Zhang, Q. Wu, Y. Liu, and O. V. Yazyev, "Magnetoresistance from Fermi surface topology". *Physical Review B* **99** (2019) 35142.

[45] A. Wang, D. Graf, Y. Liu, Q. Du, J. Zheng, H. Lei, and C. Petrovic, "Large magnetoresistance in the type-II Weyl semimetal $WP_2$". *Phys. Rev. B* **96** (2017) 121107(R).

[46] G. Autes, D. Gresch, M. Troyer, A. A. Soluyanov, and O. V. Yazyev, "Robust Type-II Weyl Semimetal Phase in Transition Metal Diphosphides $XP_2$ (X=Mo,W)". *Phys. Rev. Lett.* **117** (2016) 066402.

[47] M. Pi, T. Wu, D. Zhang, S. Chen and S. Wang, "Phase-controlled synthesis and comparative study of *α*- and *β*-WP$_2$ submicron particles as efficient electrocatalysts for hydrogen evolution". *Electrochimica Acta* **216** (2016) 304–31.

[48] F. C. Chen, H. Y. Lv, X. Luo, W. J. Lu, Q. L. Pei, G. T. Lin, Y. Y. Han, X. B. Zhu, W. H. Song, and Y. P. Sun, "Extremely large magnetoresistance in the type-II Weyl semimetal MoTe$_2$". *Physical Review B* **94** (2016) 235154.

[49] D. V. Khveshchenko, "Magnetic-Field-Induced Insulating Behavior in Highly Oriented Pyrolitic Graphite". *Phys. Rev. Lett.* **87** (2001) 206401.

[50] X. Du, S.-W. Tsai, D. L. Maslov, and A. F. Hebard, "Metal-Insulator-Like Behavior in Semimetallic Bismuth and Graphite". *Phys. Rev. Lett*. **94** (2005) 166601.

[51] Y. Kopelevich, J C M Pantoja, R R da Silva and S Moehlecke, "Universal magnetic-field-driven metal-insulator-metal transformations in graphite and bismuth". *Phys. Rev. B* **73** (2006) 165218.

[52] S. Wolgast et al., Low-temperature surface conduction in the Kondo insulator SmB$_6$". *Phys. Rev. B*. **88** (1988) 180405.

[53] J. Wang, L. Li, W. You, T. Wang, C. Cao, J. Dai and Y. Li, "Magnetoresistance and robust resistivity plateau in MoAs$_2$". *Scientific Reports* **7** (2017) 15669.

[54] R. Singha, A. Pariari, B. Satpati, and P. Mandal, "Magnetotransport properties and evidence of a topological insulating state in LaSbTe". *Physical Review B* **96** (2017) 245138.

[55] Y. L. Wang, L. R. Thoutam, Z. L. Xiao, J. Hu, S. Das, Z. Q. Mao, J. Wei, R. Divan, A. Luican-Mayer, G. W. Crabtree and W. K. Kwok, "Origin of the turn-on temperature behavior in WTe$_2$". *Physical Review B* **92** (2015) 180402.

[56] A. B. Pippard, "Magnetoresistance in Metals" In.: Cambridge University Press, 1989.

[57] R. H. McKenzie, J. S. Qualls, S. Y. Han and J. S. Brooks, "Violation of Kohler's rule by the magnetoresistance of a quasi-two-dimensional organic metal". *Physical Review B* **57** (1998) 11854–11857.

[58] Z. J. Yue, X. L. Wang and S. S. Yan, "Semimetal-semiconductor transition and giant linear magnetoresistances in three-dimensional Dirac semimetal Bi$_{0.96}$Sb$_{0.04}$ single crystals". *Appl. Phys. Lett.* **107** (2015) 112101.

[59] H. Li, H. He, H.-Z. Lu, H. Zhang, H. Liu, R. Ma, Z. Fan, S.-Q. Shen & J. Wang, "Negative magnetoresistance in Dirac semimetal $Cd_3As_2$". *Nat. Commun.* **7** (2016) 10301.

[60] M. Novak, S. Sasaki, K. Segawa, and Y. Ando, "Large linear magnetoresistance in the Dirac semimetal TlBiSSe". *Phys. Rev. B* **91** (2015) 41203.

[61] K. Takiguchi, Y. K. Wakabayashi, H. Irie, Y. Krockenberger, T. Otsuka, H. Sawada, S. A. Nikolaev, H. Das, M. Tanaka, Y. Taniyasu and H. Yamamoto, "Quantum transport evidence of Weyl fermions in an epitaxial ferromagnetic oxide", *Nat. Comm.* **11** (2020) 4969.

[62] M. M. Parish and P. B. Littlewood, "Non-saturating magnetoresistance in heavily disordered semiconductors". *Nature* **426** (2003) 162–165.

[63] D. Miller, K. Kubista, G. Rutter, M. Ruan, W. de Heer, P. First, and J. Stroscio, "Observing the Quantization of Zero Mass Carriers in Graphene". *Science* **324** (2009) 924–927.

**FIGURE CAPTIONS**

**Fig. 1:** The Reitveld refined X-ray diffraction pattern of sample. Left inset shows the EDX spectrum of polycrystalline $WP_2$. Right inset shows the crystal structure of $WP_2$. W and P atoms are represented by dark red and blue colours, respectively.

**Fig. 2:** (a) The zero field resistivity $\rho_{xx}(T)$ of a sample as a function of temperature. Solid line represents a fit to equation: $\rho_{xx}(T) = \rho_{xx}(0) + BT^n$. Inset shows the plot of resistivity variation as $T^3$ at low temperatures. (b) The resistivity at 0T is fitted with an equation: $(T) = \rho_0 + aT^2 + bT^5 + c^*exp(-T_0/T)$ below 100K (red line) and linear fitted above 250K (green line).

**Fig. 3:** (a) The temperature dependent resistivity taken at different applied magnetic fields. Inset shows the enlarged view of the resistivity at all fields in the low temperature region. (b) The plot of first order of temperature derivative of resistivity $d\rho/dT$ versus temperature at fixed magnetic fields. The plot of $\ln \rho$ versus $T$ at 7T and 11T applied magnetic field. (d) The temperature dependence of magnetoresistance $MR(T)(\%) = (\rho(T,B) - \rho(T,0))/\rho(0T) \times 100\%$ under different applied fields. Inset depicts the normalized i.e., MR/MR(2K) at constant fields.

**Fig. 4:** (a) The transverse magnetoresistance $MR(\%) = (R(B) - R(B = 0))/R(B = 0) \times 100\%$ as a function of magnetic field at different temperatures. (b) Kohler scaling of field dependent MR at distinct temperatures. (c) MR versus $B/\rho_0$ fitted using power law at different temperatures using Kohler's rule in Equation (1). (d) The temperature variation of the parameters $\alpha$ and $m$ derived from the Kohler fitting of MR at all temperatures.

**Fig. 5:** The magnetic field dependence of MR measured at 2K and 5K is fitted with an equation: $MR = a_1B + a_2B^2$. (b) The field derivative of MR i.e., $dMR/dB$ plotted against the magnetic field at 2K and 5K. Inset illustrates the plot of $dMR/dB$ vs. B in the high field region.

**Fig. 6:** Temperature dependence of resistivity at 0T and 11T and their difference $\Delta\rho_{xx}$. The solid lines are fit to Equation (2) with $\alpha = 0.565 \ (\mu\Omega - cm/T)^{1.54}$ and $m = 1.54$.

**Fig. 1**

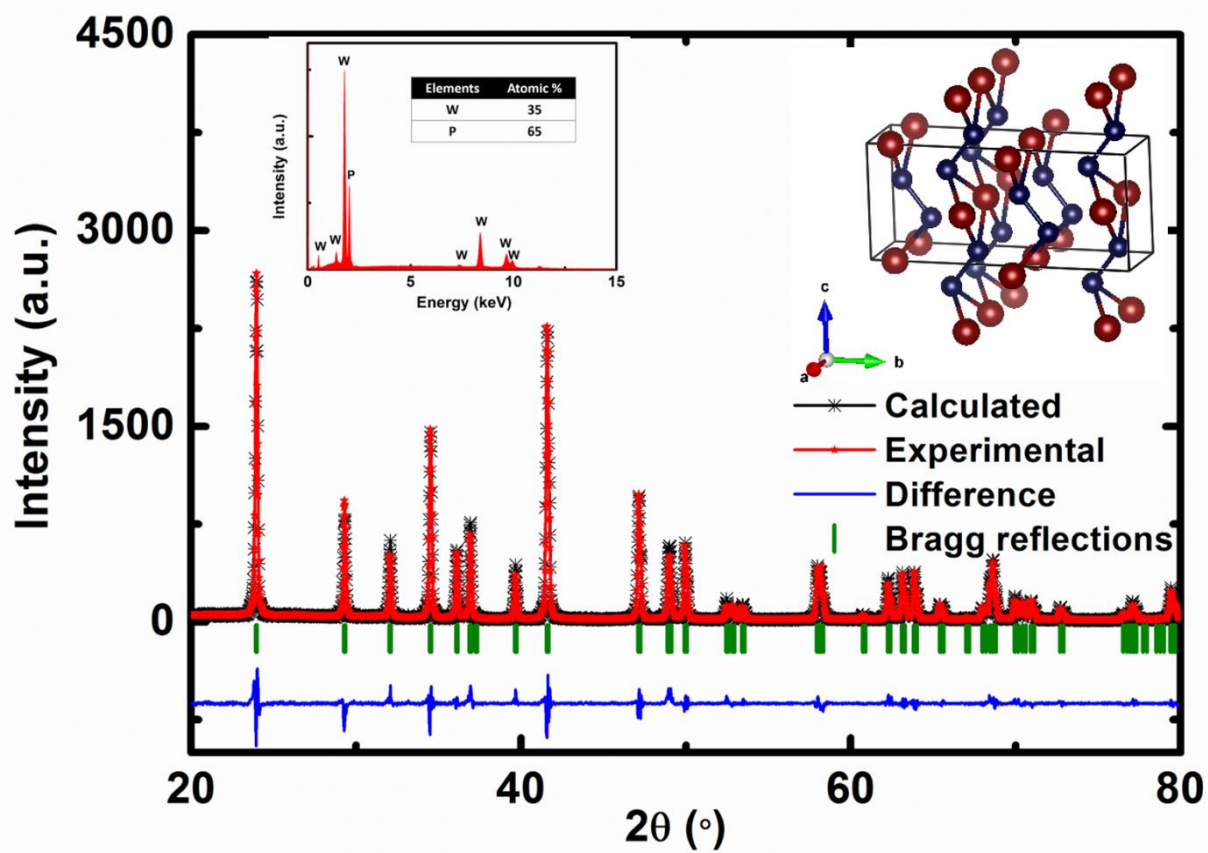

**Fig. 2**

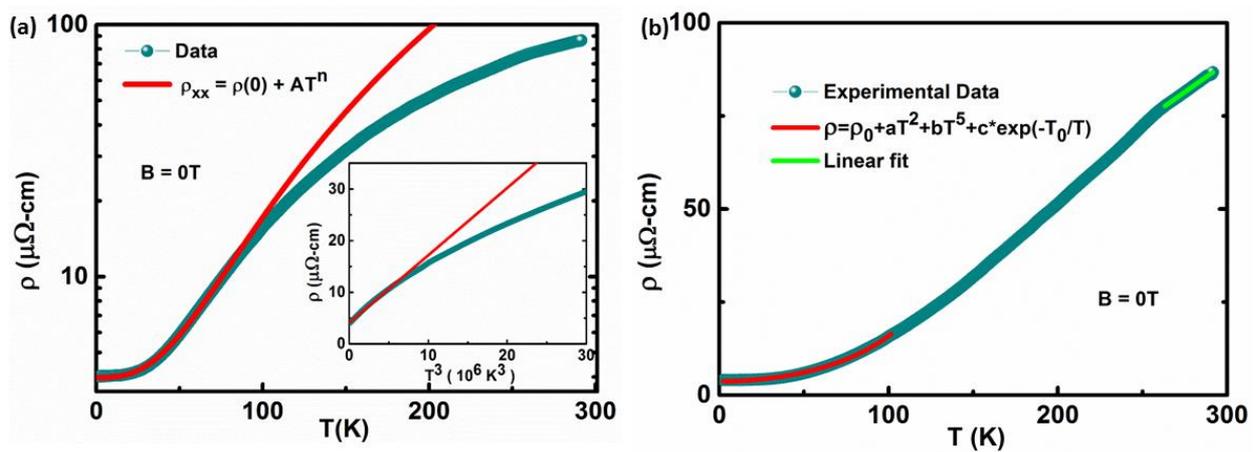

**Fig. 3**

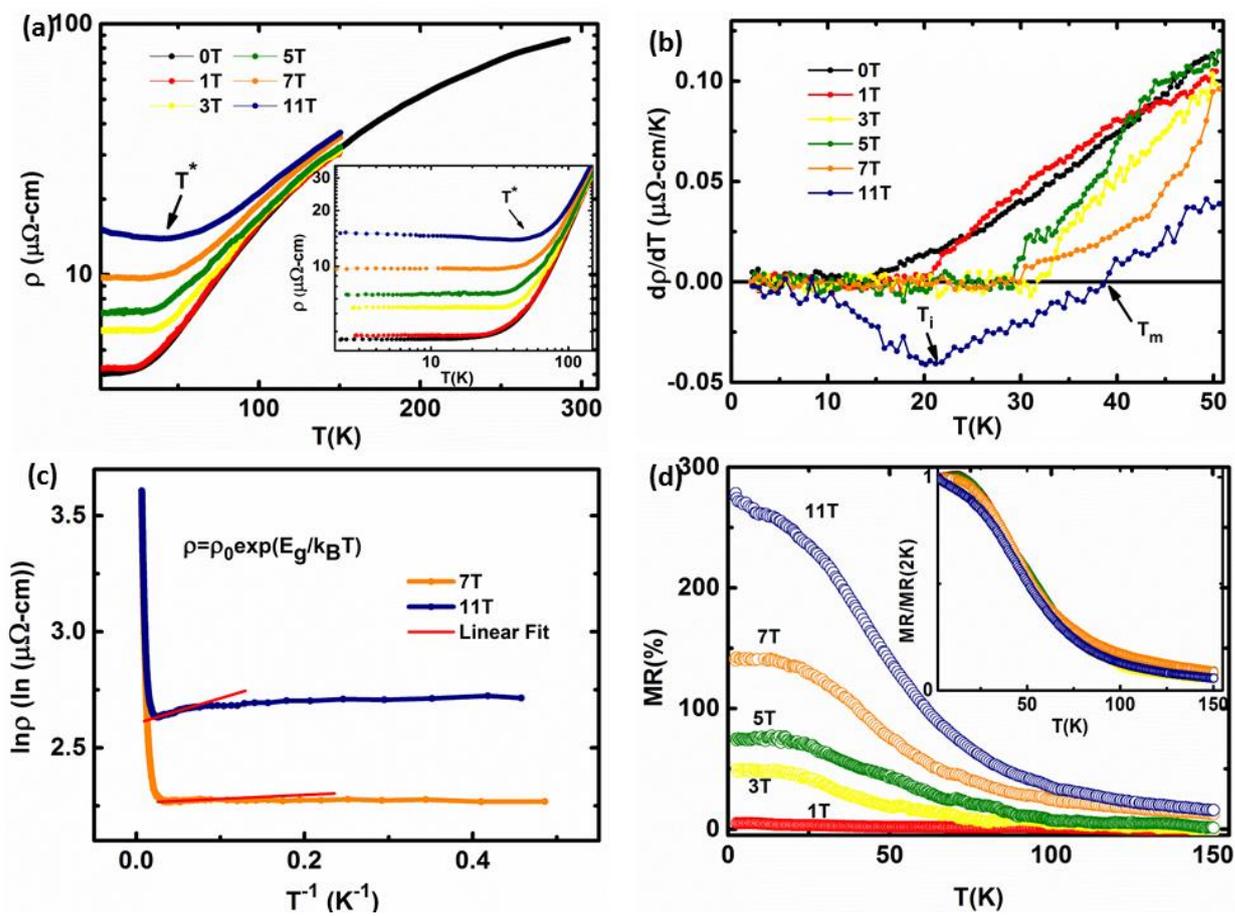

**Fig. 4**

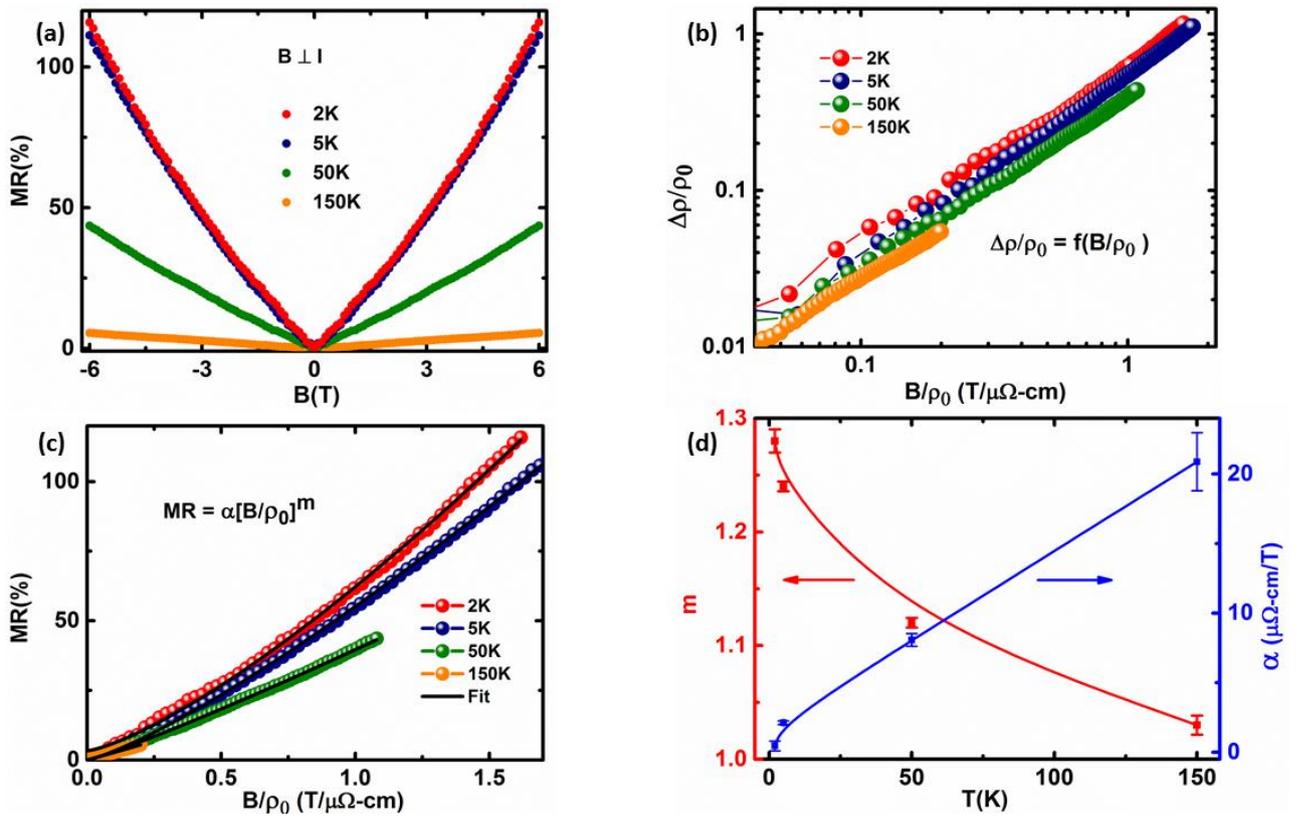

**Fig. 5**

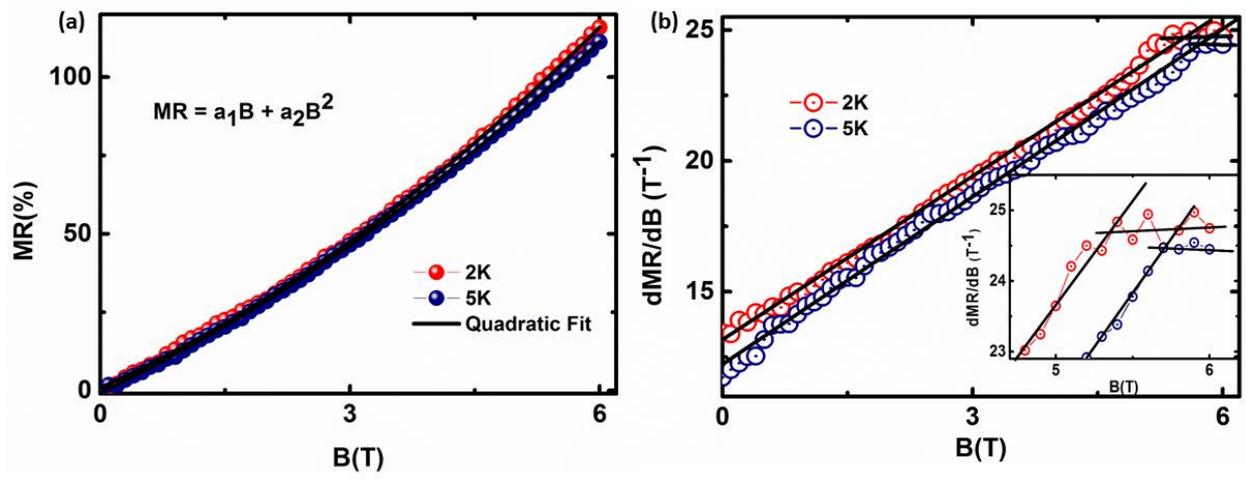

**Fig. 6**

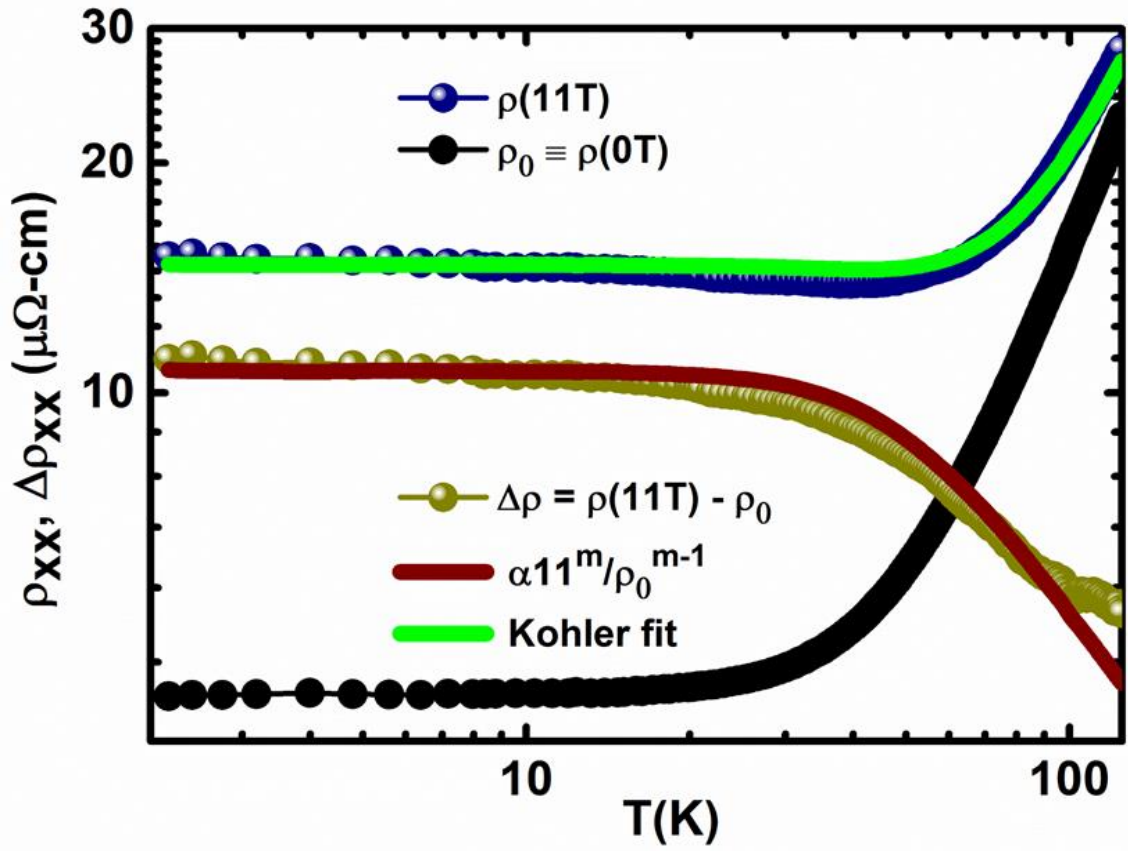